\documentclass[twocolumn,aps,prl,showpacs,superscriptaddress]{revtex4}
\usepackage{amssymb}
\usepackage{amsmath}
\usepackage{graphicx}
\usepackage{epsfig}
\begin{document}

\title{Quest for quantum superpositions of a mirror: high and moderately low temperatures}
\author{J\'ozsef Zsolt Bern\'ad}
\email{bernad@complex.elte.hu}
\affiliation{Department of Physics of Complex Systems, E\"otv\"os University, Budapest, Hungary}
\author{Lajos Di\'osi} 
\email{diosi@rmki.kfki.hu}
\affiliation{Research Institute for Particle and Nuclear Physics, Budapest, Hungary}
\author{Tam\'as Geszti}
\email{geszti@complex.elte.hu}
\affiliation{Department of Physics of Complex Systems, E\"otv\"os University, Budapest, Hungary}

\begin{abstract}
The Born-Markov master equation analysis of the vibrating mirror and photon experiment proposed 
by Marshall, Simon, Penrose and Bouwmeester is completed by including the important issues of
temperature and friction. We find that at the level of cooling available to date, visibility revivals
are purely classical, and no quantum effect can be detected by the setup, no matter how strong 
the photon-mirror coupling is. Checking proposals of universal nonenvironmental decoherence 
is ruled out by dominating thermal decoherence; a conjectured coordinate-diffusion contribution 
to decoherence may become observable on reaching moderately low temperatures.
\end{abstract} 
\maketitle

The nature of the quantum-classical border along the mass scale is still poorly defined. There remain some 
10 orders of magnitude unexplored between the heaviest molecules for which c.o.m. interference has 
been observed \cite{zeil}, and the lightest nanomechanical objects, for which no quantum behavior has been 
seen \cite{mirrexp}. In trying to close the gap top down, the primary experimental task is to find firm 
evidence, never seen so far, that the spatial motion of a mass as large as a nanomechanical object does 
follow the Schr\"odinger equation, notwithstanding environmental interactions, or noise, which would quickly
decohere the wave function. Only having succeeded in suppressing that effect so that interference of a 
heavy object is detected beyond any doubt, can we turn to checking the presence of spontaneous (also called 
universal or intrinsic) decoherence \cite{non-env} on top of the environmental one. 

An experimentally accessible system with potentialities to achieve the above goal is a photon in a 
high-quality resonating cavity, coupled by its radiation pressure to a nanomechanical oscillator, 
carrying one of the mirrors that close the cavity. After pioneering experiments \cite{mirrexp} which 
did not detect any quantum effect on the mirror, as well as thoughtful theoretical analyses 
\cite{law}, a promising idea appeared for bridging the frequency gap and carrying out a genuine quantum 
test \cite{marshall,others}. In that proposal, the vibrating mirror closes an optical cavity in arm $A$ 
of a Michelson interferometer, arm $B$ having another cavity with fixed mirrors. The vibrating mirror is 
expected to become entangled with a single photon traveling along both arms, the mirror 
being split into a kind of Schr\"odinger cat doublet. The interference of the photon is detected with 
the scope of extracting information about the quantum motion of the mirror. Since the vibrations of 
the mirror are much slower than the frequency of light, a shift of the interference pattern would be 
unobservable; the good chance is to record the {\em visibility} which is modulated by the motion of 
the mirror, creating revivals as the components of the superposition overlap again and again. 

Highly worth doing as it is, this is a very hard experiment, for various reasons. One 
thing is that high-quality optical resonators are needed to keep the photon alive for
several, or at least one, return of the mirror; a less familiar task is to preserve coherence
of the vibrating mirror itself. The latter requires efficient cooling and a drastic reduction
of various mechanisms of environment-induced decoherence, at least partly related to friction.

Drawing on previous analyses of the vibrating mirror and photon problem, the experimental proposal 
\cite{marshall} has been analyzed by Adler {\it et al.} \cite{adler}, bringing considerable new insight 
into the way decoherence influences the interferometric signal. The present Letter is meant to add 
the crucial features of finite temperature and friction, which have been but qualitatively described in 
Ref. [\onlinecite{marshall}]. The main point is that the requirement of sufficiently strong coupling to 
create entanglement enforces the use of a low-frequency vibrating mirror. Then, however, unless cooling 
performances are considerably improved, one remains in the high-temperature range, where no genuine 
quantum effect can be observed. We also confirm that testing theoretical proposals about universal 
nonenvironmental decoherence mechanisms has remained an extremely bold enterprise for some time to come.
The only quantum effect accessible on moderate progress in cooling would be a refinement of the 
treatment of quantum friction, proposed in Ref. [\onlinecite{dllind}], see below.

We start out in the framework posed by Adler {\it et al.}, calculating visibility of the photon 
interference as $\nu(t)=2|{\rm Tr}_m\hat\rho_{OD}(t)|,$ where $\hat\rho_{OD}=
\,_A\langle1|_B\langle0|\hat\rho|0\rangle_A|1\rangle_B$ is the off-diagonal element of the full density 
matrix $\hat\rho$ of the mirror-photon system, where $|0\rangle_j$ and $|1\rangle_j$ are photon states
with 0 or 1 photon, respectively, in arm $j=A,B$. 

For $\hat\rho_{OD},$ Adler {\it et al.} \cite{adler} derive a master equation, with a position decoherence 
term of strength $D_{pp}$. To include friction of constant $\gamma$ in the treatment, we follow the usual 
theoretical pathway using the high-temperature Markovian master equation \cite{caldleg} of quantum friction. 
Following Ref. [\onlinecite{dllind}], we also include a correction term, negligible for high temperatures 
but relevant for moderately low ones, taking the form of a momentum decoherence term of strength 
$D_{qq}$; we shall tune it toward its theoretical minimum value $\gamma^2/16D_{pp}$ assuring to preserve 
positivity of the density matrix \cite{dllind}. This correction, as we shall see later, may turn out to 
govern the only quantum effect detectable at moderately low temperatures.

With all those extra terms, using units with $\hbar=1$, the master equation  
reads 
\begin{equation}
\label{1} \begin{split}
\frac{\partial \hat\rho_{OD}}{ \partial t}
=&-i\hat H^A \hat\rho_{OD}+i\hat\rho_{OD} \hat H^B - D_{pp}
[\hat x,[\hat x,\hat\rho_{OD}]]\\&
-i\frac{\gamma}{2}[\hat x,\{\hat p, \hat\rho_{OD}\}]
- D_{qq}[\hat p,[\hat p, \hat\rho_{OD}]],
\end{split}
\end{equation}
with
\begin{equation}
\hat H^B =\frac{M\omega_m^2}{2} \hat x^2 + \frac{1}{2M} \hat p^2, \,\,\,\,\,\,\,\,\,\,  
\hat H^A = \hat H^B - \omega_c\frac{\hat x}{L},
\end{equation}
where $\omega_m$ is the frequency of the vibrating mirror \cite{law}, 
$L$ the cavity length, $\omega_c$ the frequency of the photon, and $M$ the mass of the mirror.

Besides using the width $\sigma=1/\sqrt{2M\omega_m}$ of the ground-state wave packet of the mirror 
and the oscillator quality factor $Q_m^{-1}=\gamma/\omega_m$, the following dimensionless parameters
will be of central importance for the discussion below: the photon-mirror coupling constant 
$\kappa=(\omega_c/\omega_m)(\sigma/L)$ \cite{law}, the decoherence strength 
$\Lambda=(\sigma^2/\omega_m)D_{pp}$, and the combination $\chi=D_{qq}/(\omega_m\sigma^2)$. 
Using these notations, and temporarilly introducing units of time with $\omega_m=1,$ 
the final form of the master equation becomes
\begin{equation}
\begin{split}
\label{4}
\frac{\partial \hat\rho_{OD}}{ \partial t}
=&-i\sigma^2[\hat p^2, \hat\rho_{OD}]-\frac{i}{4}\sigma^{-2}[\hat x^2, \hat\rho_{OD}]
+i\kappa\, \sigma^{-1}\,\hat x\,  \hat\rho_{OD}\\&
-\Lambda\sigma^{-2}[\hat x,[\hat x,\hat\rho_{OD}]]
-\frac{i}{2}\,Q_m^{-1}[\hat x,\{\hat p,\hat\rho_{OD}\}]\\&
-\chi\sigma^2\,[\hat p,[\hat p,\hat\rho_{OD}]].
\end{split} 
\end{equation}
The above equation can be solved analytically, {\sl e. g.} via the trace expression \cite{zurgiulini}:
\begin{equation}\label{trace}
\tilde\rho_{OD}(k, \Delta)= 
       {\rm Tr}_m\left( \hat\rho_{OD} \,\, \exp\, i(\sigma^{-1} k\hat x+\sigma \Delta\hat p)\right),
\end{equation}    
where $k,\,\Delta$ are dimensionless Fourier variables \cite{diododd}. Using this representation, the 
master Eq. (\ref{4}) results in the following equation of motion:
\begin{equation}
\begin{split}
\label{6}
& \frac{\partial \tilde\rho_{OD}(k, \Delta)}{ \partial t} 
= 2\,k \frac{\partial}{\partial \Delta} \tilde\rho_{OD}
- \frac12 \Delta \frac{\partial}{\partial k}\tilde\rho_{OD}
- \Lambda \Delta^2 \tilde\rho_{OD}\\&
\,\,\,\,\,+\kappa \left(\frac{\partial}{\partial k}+i\,\frac{ \Delta}{2}\right)\tilde\rho_{OD} 
-Q_m^{-1} \Delta \frac{\partial}{\partial \Delta} \tilde\rho_{OD} 
- \chi k^2 \tilde\rho_{OD}. 
\end{split} 
\end{equation}

We aim at finding a temperature-averaged solution. Following the tradition \cite{law}, we first  
solve the equations for an arbitrary pure coherent state $| \alpha_0 \rangle$ of the 
mirror, for which Eq. (\ref{trace}) takes the 
Gaussian form
\begin{equation}\label{expci}
\tilde\rho_{OD}(k, \Delta)=\frac12{\rm e}^{-[c_1k^2+c_2k \Delta+c_3 \Delta^2+ic_4k+ic_5 \Delta +c_6]},
\end{equation}
with the initial values
\begin{equation}
\begin{split}
c_1(0)&=\frac12,\,\,\,\,c_2(0)=0,
\,\,\,\,c_3(0)=\frac18, \\
c_4(0)&=-2Re[\alpha_0],\,\,\,
c_5(0)=-Im[\alpha_0],\,\,
c_6(0)=0.
\end{split} 
\end{equation}
The corresponding visibility is $2{\rm Tr}_m \hat\rho_{OD}(t,\alpha_0)=e^{-c_6(t)}$, to be evaluated
using Equation (\ref{6}) which preserves the Gaussian structure (\ref{expci}) in time, 
with coefficients evolving according to the following simple linear equations: 
\begin{equation}
\begin{split}
\label{8}
\dot{c}_1&=2c_2+\chi,\,\,\,\,
\dot{c}_2=4 c_3- c_1-Q_m^{-1} c_2,\\
\dot{c}_3&=-\frac12 c_2 -2 Q_m^{-1} c_3 +\Lambda,\,\,\,\,
\dot{c}_4=2 c_5-2i \kappa \, c_1,\\
\dot{c}_5&=-\frac12 c_4 - \kappa\left(ic_2+\frac12\right)-Q_m^{-1} c_5, \,\,\,\,
\dot{c}_6=i\kappa\,c_4.
\end{split} 
\end{equation}
The solution depends on $\alpha_0$ in the form 
$c_6(t)=\kappa^2\, f_1(t)-i\kappa\left\{{\rm Re}[\alpha_0]f_2(t)
 -{\rm Im}[\alpha_0]f_3(t)\right\}$;
then we do the thermal averaging of $e^{-c_6(t)}$ over $\alpha_0$  \cite{scully} to obtain
\begin{equation}\begin{split}
\nu(t)&=2\left|\int P_T(\alpha_0)\, 
{\rm Tr}_m\, \hat\rho_{OD}(t;\alpha_0)\, d^2 \alpha_0\right|\\&
=\left|e^{-\kappa^2\left[\,f_1(t)
        +\frac{\bar n}{4}\left(f_2^2(t)+f_3^2(t)\right)\right]}\right|
\end{split}\end{equation}
where $P_T(\alpha_0)=e^{-|\alpha_0|^2/\bar n}/(\pi\bar n)$ is the P function of the initial 
 thermal equilibrium state of the mirror, with 
$\bar n=\left[\exp(\hbar \omega_m/k_BT)-1)\right]^{-1}$. 
\noindent The functions $f_1,\,f_2,\,f_3$ are obtained in analytical form; however, the resulting 
formulas are not transparent enough to be displayed in full generality. Simpler results are obtained 
for the relevant case of a high-quality mechanical oscillator: $Q_m^{-1}\ll 1$. Then, while evaluating 
the complex frequencies in full accuracy, in the amplitudes one has to keep only the leading-order 
corrections in $Q_m^{-1}$. With that simplification, returning to physical units of time and introducing 
$\tilde\omega_m=\sqrt{\omega_m^2-(\gamma/2)^2}$ which is the frequency of the damped classical 
oscillator, we arrive at our final result 
\begin{equation}
\begin{split}\label{simple}
&\nu(t)=\exp\left\{-\left(\bar n +1/2\right)\kappa^2\left[
 1+e^{-\gamma t}-2e^{-\frac{\gamma}{2}t}\cos(\tilde\omega_mt)\right]\right\}
\\&\,\,\,\,\times\exp\left(-6\kappa^2\Lambda\left\{ \tilde\omega_m t
         \left[\frac{1-e^{-\gamma t}}{3\gamma t}
                       \left(1+\frac{\chi}{4\Lambda}\right)+\frac{2}{3}\right]
                        \right. \right. \\ & \left.\left.
 \,\,\,\, -\frac{4}{3} e^{-\frac{\gamma}{2}t}\sin(\tilde\omega_m t)
                        \right. \right. \\ & \left.\left.
 \,\,\,\, +\,\frac{1}{6} e^{-\gamma t}\sin(2\tilde\omega_m t)\left(1-\frac{\chi}{4\Lambda}\right)
\right\}\right).
\end{split}\end{equation}

In the first of the two factors above it is easy to recognize the visibility revival effect as 
originally proposed by Marshall {\em et al.} \cite{marshall}, modified by  the temperature averaging already 
discussed by Bose {\it et al.} \cite{law}, as well as the mechanical effect of friction. The second factor
describes decoherence effects, in accordance with the result of Adler {\it et al.} \cite{adler}, now
including the coordinate diffusion correction $\chi$ \cite{dllind} (see below), also modified by friction. 

Inference about decoherence mechanisms can be extracted from the hight of the {\sl first} 
revival at $t_1=2\pi/\tilde\omega_m$: for times as short as that, damping through mechanical friction 
can be fully neglected, and (\ref{simple}) simplifies to 
\begin{equation}\label{sssimple}
\nu(t_1)=\exp\left\{-\pi\kappa^2(12\Lambda+\chi)\right\}
\end{equation} 
(see Fig. 1). We postpone the discussion of $\chi$ which is negligible at present-day temperatures (see 
below), and write tentatively $\Lambda=\Lambda_T+\Lambda_{nonenv}$. Concerning the first, dominant term, 
in thermal environment classical friction is always accompanied by classical momentum diffusion of 
strength $D_{pp}^T=Mk_BT\gamma$. This mechanism survives for quantum friction as well, causing 
the familiar thermal position decoherence 
\begin{equation}\label{CL}
\Lambda_T=(k_B\;T/2\hbar\omega_m)Q_m^{-1},
\end{equation}
where we have restored the true physical scale of $\hbar$. 
Substituting the Marshall {\sl et al.} figures \cite{marshall} about present-day possibilities, 
$\omega_m=3\times10^3 s^{-1}$, $T=2\times 10^{-3} K$, and $Q_m=10^5$, we obtain $\Lambda_T\approx 0.5$.
That is the background against which non-environmental decoherence mechanisms expected from the models 
Ghirardi-Rimini-Weber, ``Quantum Mechanics with Universal Position Localization'' (QMUPL) or ``Continuous
Spontaneous Localization'' (CSL) \cite{non-env} should be tested, according to the suggestion of 
Marshall {\it et al.} The estimates range from  $\Lambda_{CSL}\approx 0.2\times 10^{-8}$, to much 
smaller figures for gravitation-related universal collapse model \cite{DioPen}. This indicates that, 
for the thermal background, many orders of magnitude should be gained in friction and cooling before 
nonenvironmental decoherence proposals might show up in the experiment.

   We must not forget our basic task, to see if the proposal \cite{marshall} can yield 
evidence at least for the quantum behavior of the vibrating mirror. At the temperatures of mK and 
vibration frequencies of kHz envisaged for the mirror, we have $\bar n\sim10^5$ which means the mirror 
is well in its high-temperature regime; accordingly, quantum effects are expected to be masked altogether,
which is confirmed by the parameter combinations appearing in Eq.~(\ref{simple}). Indeed, the 
parameter $\chi$ can be ignored at high temperatures, and the other two relevant combinations 
\begin{equation}\label{visext}
\kappa^2\Lambda_T=\frac{k_BT}{4M\omega_m^2 L^2}\frac{\omega_c^2}{\omega_m^2}Q_m^{-1},
\,\,\,\,\, \kappa^2\bar n=\frac{k_BT\omega_c^2}{2M\omega_m^4 L^2}, 
\end{equation}
turn out to be fully classical, containing no factor of $\hbar$. The first of them is the visibility 
extinction coefficient (cf. the similar result by Bose {\it et al.} \cite{law}); the second is the parameter 
that controls thermal narrowing of the duration of visibility recurrences.
Since $\kappa^2\bar n\gg1$ at high temperatures, the duration of visibility revivals will be much shorter 
than the vibration period. That temperature-related narrowing effect has been already mentioned by Marshall 
{\it et al.} \cite{marshall} as a challenge to the stability of the experimental setup.

The existence of visibility revivals in no way contradicts to the full classicality manifest in our
results. Indeed, mirror-photon entanglement and cyclically returning disentanglements {\sl coincide} 
with classical mirror-light correlation and returning decorrelations (classical radiation being scaled 
to one-photon strength), as the mirror repeatedly passes through its initial position, {\sl independently 
of initial conditions}. That robust periodicity is specific to harmonic oscillator dynamics. All that can 
be followed in detail through the appropriate equations \cite{class}. 

\begin{figure}[hbt]
\resizebox{60mm}{!}{
\includegraphics{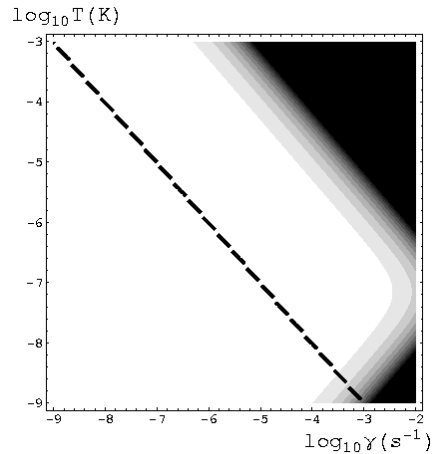}}
\caption{\label{contour} Contours of equal visibility (white for good, black for poor), as a function of 
temperature $T$ and friction $\gamma$; based on Eqs. (\ref{sssimple}), (\ref{CL}), (\ref{chi}). 
Upper-right corner, present-state possibilities; dashed line, where CLS decoherence becomes observable. 
The turnback of the contours in the lower-right corner corresponds to the coordinate diffusion contribution 
\cite{dllind}, with $\kappa=1$ and $\lambda=1$. }
\end{figure}


In order to detect quantum behavior of the mirror, we must cool it to medium temperatures where
$\bar n$ is a smaller number, say $5$ or $10$. This could be done with GHz oscillators, see e.g. 
\cite{GHz}. Our frequency cannot be that high though: a hard, high-frequency oscillator resists to 
the push the photon exerts on the mirror, as expressed in the low value of the coupling parameter 
$\kappa$. If $\kappa<1$, the push is not strong enough to split the mirror into a well-separated 
superposition cat doublet, and no entanglement is created. This strong-coupling requirement forces 
the photon-mirror system into a vicious circle: to obtain a quantum effect, one must use a relatively 
soft (low-frequency) vibrating mirror, which is hard to cool down close to its ground state; therefore 
it is hard to leave the high-temperature range.

If we still succeed in pushing temperature down to $k_BT/\hbar\omega_m\approx{\cal O}(10)$, which 
for the soft oscillator envisaged by Marshall {\it et al.} corresponds to the range of a few $\mu K$, 
the quantum correction proportional to $D_{qq}$ \cite{dllind} enters the dynamics (\ref{1}) 
and may become accessible to measurement, as discussed already by Jacobs {\it et al.} \cite{law}. 
Let us tune $D_{qq}$ toward its theoretical minimum; {\it i.e.}, we assume 
$D_{qq}=\lambda\gamma^2/16D_{pp}^T$ where $\lambda\gtrsim 1$ is a small number to be extracted 
from experiment. Evaluating the factor $1+\chi/4\Lambda_T$ in Eq.(\ref{simple}), we get the 
following medium-temperature quantum correction to the classical visibility extinction 
coefficient (\ref{visext}):
\begin{equation}\label{chi}
\frac{k_BT}{4\,M\,\omega_m^2\,L^2}\frac{\omega_c^2}{\omega_m^2}Q_m^{-1}
\left[1+\lambda(\hbar\omega_m/4k_BT)^2\right],
\end{equation}
which may reach measurability at moderately low temperatures. Approaching even that moderately 
low-temperature range is a bold enterprise with the photon-mirror combination. 
Refinement of the theoretical treatment for low temperatures is also desirable.

In summary, by including friction and temperature averaging in the theoretical framework
set by Adler {\it et al.} \cite{adler}, we gave an overall theoretical analysis of the 
experimental setup proposed by Marshall {\it et al.} \cite{marshall}. We find that
although photon visibility revivals are expected to be detected in the proposed setup, at the 
cooling level currently available they do not allow one to conclude that a macroscopic body might 
exhibit genuine quantum behavior. In agreement with the conclusion of Adler {\it et al.} \cite{adler}, 
we also confirm that detection of any of the hypothetical nonenvironmental decoherence mechanisms 
is a remote scope, being orders of magnitude weaker than present-day thermal background decoherence. 
Nevertheless, on reaching moderately low temperatures, there is the chance to detect a different 
quantum effect: a coordinate-diffusion-related contribution to decoherence. Anyway, unprecedented 
progress in cooling a soft mirror is the only obvious way towards seeing both robust quantum 
effects and eventual violation of standard quantum mechanics, which is an aim of extreme importance.

This work was partially supported by the Hungarian Scientific Research Fund OTKA
under Grant No. T049384. We are indebted to A. Bodor for many helpful discussions.

\end{document}